\documentclass{article}
\usepackage{sao2}
\usepackage{latexsym,mathtext,amssymb}
\usepackage{float,epsfig}
\usepackage[figuresright]{rotating}
\newcommand{\msun}{M$_{\odot}$}
\newcommand{\ergl}{ergs~s$^{-1}$}

\newcommand{\ergf}{ergs~cm$^{-2}$~s$^{-1}$}

\newcommand{\hei}{He~{\sc i}}
\newcommand{\heii}{He~{\sc ii}}
\newcommand{\oiii}{[O~{\sc iii}]}
\newcommand{\oii}{[O~{\sc ii}]}
\newcommand{\oi}{[O~{\sc i}]}
\newcommand{\sii}{[S~{\sc ii}]}
\newcommand{\siii}{[S~{\sc iii}]}
\newcommand{\feiii}{[Fe~{\sc iii}]}
\newcommand{\hii}{H~{\sc ii}}
\newcommand{\nii}{[N~{\sc ii}]}
\renewcommand{\ni}{[N~{\sc i}]}

\newcommand{\kms}{$km\,s^{-1}\,$}
\newcommand{\etal}{et al.}
\newcommand{\otoh}{$12+\lg(O/H)$}
\newcommand{\Ry}{\rm Ry \/}

\renewcommand{\mag}[1]{^{\rm m}\!\!\!#1\,}
\newcommand{\AAA}{  \AA \/}

\newcommand{\apjl}{\apj Letters}

\begin{document}

\title{Spectroscopy of optical counterparts of ultraluminous X-ray sources}
%%%\title{Spectroscopy of ULX Optical Counterparts on the SAO 6m Telescope}

\author{ Abolmasov P., Fabrika S., Sholukhova O., Afanasiev V.}

\institute{\saoname}
\maketitle

\begin{abstract}
Here we present the results of panoramic and long-slit observations of eight ULX nebular counterparts held with 
the 6m SAO telescope. In two ULXNe we detected for the first time signatures of high 
excitation (\oiii$\lambda$5007 / H$\beta$ $>$ 5). Two of the ULXs were identified with young 
($T \sim 5-10~Myr$) massive star clusters.
Four of the eight ULX Nebulae (ULXNe) show bright high-excitation lines.
This requires existence of luminous ($\sim 10^{38}\div 10^{40}$\ergl) UV/EUV sources 
coinciding with the X-ray sources.
Other 4 ULXNe require shock excitation of the gas with shock velocities of 20-100\kms.
However, all the studied ULXN spectra show signatures of shock excitation, but
even those ULXNe where the shocks are prevailing show presence of a hard ionizing source
with the luminosity at least $\sim 10^{38}$\ergl.
Most likely shock waves, X-ray and EUV ionization act simultaneously in
all the ULXNe, but they may be roughly separated in two groups, shock-dominated and
photoionization-dominated ULXNe. The ULXs have to produce strong winds
and/or jets powering their nebulae with $\sim 10^{39}$\ergl. Both the wind/jet activity 
and the EUV source needed are consistent with the suggestion that ULXs are high-mass X-ray 
binaries with the supercritical accretion disks of the SS433 type.
\end{abstract}
\maketitle

%%%%%%%%%%%%%%%%%%%%%%%%%%%%%%%%%%%%%%%%%%%%%%%%%%%%%%%%%%%%%%%%%%%%%%%%%%%%%
\section{Introduction}
\subsection{Ultraluminous X-ray Sources}

A point-like X-ray source is considered an Ultraluminous X-ray Source (ULX) if its luminosity 
exceeds $10^{39} erg\,s^{-1}$ and it does not coincide with the galactic nucleus (to 
exclude active galactic nuclei). 
%It also takes sense to exclude young (such as several years) Supernova Remnants (SNR), which can shine as bright as $10^{41} erg\,s^{-1}$, like the remnant of SN1988Z (Fabian \& Terlevich, 1996). 
There are more than 150 ULXs known at the present time (Swartz, 2004) but very little is clear yet about the physical nature of these objects.

Luminosity of an accreting source is limited by the Eddington value, that for a stellar mass ($\sim 10$\msun) black hole is about $10^{39}$\ergl. 
The limit, however, can be violated by a logarythmic factor if accretion is supercritical 
(Shakura \& Sunyaev, 1973; Abramowicz \etal, 1980).
ULXs most likely are accreting black holes (their compactness follows from relatively fast 
variability of some of them (Krauss \etal, 2005; Soria \etal, 2006)), and their apparent luminosities are well 
above the Eddington limit for a stellar mass ($\sim 10$\msun) black hole. 
Different models were proposed to explain the ULX phenomenon, basing on three main ideas:
\begin{itemize}
\item[-] the black hole mass may be larger, $\sim 100-10000$ solar masses, 
-- Intermediate Mass Black Holes, IMBHs (Madau \& Rees, 2001; Colbert \& Miller, 2005)
\item[-] accretion may be supercritical but still radiatively efficient (Begelman, 2002)
\item[-] observed source may be intrinsically anisotropic (Fabrika \& Mescheryakov, 2001; King \etal, 2001)
\end{itemize}

%%%Any combination of the three effects is possible.
%%%For example, supercritical accretion disks are likely to be highly anisotropic sources \cite{FaMe01}. 
IMBHs are supposed to accrete matter from massive donor stars, accretion from 
interstellar medium is unlikely to provide enough material.
Possible observational appearences of IMBH+donor close binaries were considered by Hopman \etal~(2004) and Copperwheat \etal~(2005).
IMBHs are considered either compact remnants of Population III stars
(Madau \& Rees, 2001) or products of stellar collisions in stellar cluster or protocluster cores 
(Soria, 2006). 

Supercritical accretion disks (SCADs) are the most appreciated alternative to the scenarios involving IMBHs. 
The best candidates are massive black-hole binaries like SS433, the only example of a 
supercritical accretor in the Galaxy (Fabrika, 2004).

\subsection{Optical Counterparts} % and Environment}

For most of the well-studied ULXs optical counterparts were found. 
In some cases optical observations reveal stellar counterparts -- usually OB supergiants (Liu \etal, 2001; Terashima \etal, 2006). 
Many of ULXs are situated inside star-forming regions and therefore they are  
heavily absorbed (Terashima \etal, 2006).

Large number of ULXs are located inside nebulae (ULX nebulae, ULXNe), sometimes
superbubbles more than hundred parsecs in size (Pakull \& Mirioni, 2003; Pakull \etal, 2006),
usually having spectra typical for shock-powered nebulae (bright \sii$\lambda$6717,6731, \nii$\lambda$6748,6583, \oi$\lambda$6300,6364 and \ni$\lambda$5200 lines).
In some cases the appearence of a ULXN may be quite different, like the
high-excitation photoionized \hii\ region of HoII~X-1 (Lehmann \etal, 2005)
or a compact very bright shell-like nebula with anomalous excitation conditions
(the nebula MF16, (Abolmasov \etal 2006; Abolmasov \etal 2007a)).

One of the interesting properties of some ULXNe is \heii$\lambda$4686 emission line.
Due to this reason,
ULXNe were considered X-ray Ionized Nebulae (XINe,
Pakull \& Mirioni, 2003). This is often considered an argument for truely large X-ray 
luminosities of ULXs (Kaaret \etal, 2004). 
However, extreme ultraviolet radiation in the wavelength range $\lambda \sim 100-200\AAA$ 
has larger ability for producing \heii$\lambda$4686 recombination emission line than X-rays. Therefore
nebulae like MF16 and that around HoII~X-1 may be not X-ray Ionized Nebulae. 
For more details see discussion in Abolmasov~\etal~(2007a).
In some cases like HoIX~X-1 and NGC1313~X-1 (Gris\'e \etal, 2006; Pakull \etal, 2006 ) the HeII emission is 
broadened by $\sim 1000$\kms and can be produced either in the accretion disk or in a Wolf-Rayet donor atmosphere. 

Many of ULXs are located close to young stellar clusters, often at tens to hundreds 
parsecs from the clusters or groups of massive stars (Zezas \etal, 2002). 
There are numerous indications for the ULXs are related to young (5-10~Myr) stellar population 
(Soria \etal, 2005; Zezas \etal, 2002; Abolmasov \etal, 2007b). 
For distant ULXs ($D \gtrsim 15$~Mpc) bright star clusters with surrounding nebulae 
become the only association available for optical studies.

In the following section~\ref{sec:OBS} we describe our observations of ULX counterparts with the 6m BTA 
telescope. In section~\ref{sec:smod} results of the spectral analysis and modelling of 
the optical counterparts are presented. In section~\ref{sec:trep} we discuss the 
%%%implications for our understanding the nature of 
inference on the nature of ULXs which may follow from the observations of the ULXNe. 
%%% from observing their optical counterparts.

%%%%%%%%%%%%%%%%%%%%%%%%%%%%%%%%%%%%%%%%%%%%%%%%%%%%%%%%%%%%%%%%%%%%%%%%%%%%%
\section{Observations and Data Reduction}\label{sec:OBS}

%%%Proposing for spectral observations on the 6m we focused on the objects known to have 
We have focused on targets with known
optical counterparts, situated in nearby galaxies ($D \lesssim 10Mpc$). 
The only exception is NGC7331 X-1, 
but in that case a bright host cluster of $V \sim 20\mag{\,}$ is present.
All the host galaxies are of late Hubble types. 
In table~\ref{tab:ulxlist} we list the ULX counterparts observed. 
We used accurate Chandra X-ray coordinates ($\pm 0.3\arcsec$) according to Swartz \etal 
(2004), Roberts \etal (2003) in
the case of IC342~X-1, Terashima \& Wilson (2004) in the case of M51~X-7 and Mescheryakov (2004) for HoIX~X-1.
%%The coordinates are given according to Swartz \etal (2004) except HoIX~X-1, 
%%in case of which we use {\it ROSAT} coordinates 
%%\cite{hoix_laparola}. 

All the data were obtained with two spectrographs: Multi-Pupil Fiber Spectrograph MPFS 
(Afanasiev \etal, 2001) and SCORPIO (Afanasiev \& Moiseev, 2005) focal reducer in long-slit mode.
3D data have the advantage of providing unbiased flux calibrations. 
However, SCORPIO has higher quantum efficiency by a factor of up to $\sim 6$. 

\begin{sidewaystable*}\label{tab:ulxlist}
\caption{Observed ULXs. The columns present the target IDs, ULX X-ray coordinates (JD\,2000),
distances D to the host galaxies in Mpc,
Galactic absorption $A_V$ in the source direction according to Schlegel (1998),
angular distances R of targets towards the center of their galaxies in arcminutes,
oxygen abundances according to Pilyugin et al. (2003) (the central abundance, but interpolated
abundances to the target angular distance from the center are given in brackets),
comments on the optical counterparts, exposure time in seconds, spectral range in $\AAA$,
spectral resolution in $\AAA$, in the MPFS oservations and the same parameters in 
the SCORPIO observations.
%ULXs with bright optical counterparts observed on the SAO 6m telescope. 
%R is the angular distance towards the center of the galaxy in arcminutes.
%$A_V^{Gal}$ is the Galactic absorption in the source direction according to \cite{schlegel_abs}. 
%Oxygen abundances are given according to \cite{pilyugin}. 
%If two values are given separated by a semicolon than the first is the central 
%abundance and the second is interpolated using central abundances and gradients for the 
%actual angular distance towards the center without deprojection. 
%$[Fe/H]$ value for HoIX is taken from \cite{maketal}.}
}
\footnotesize
{ \renewcommand{\itshape}{\/}
\begin{tabular}{p{2cm}p{1.5cm}p{1.5cm}p{1cm}ccp{2.5cm}p{3cm}p{0.8cm}p{0.8cm}p{0.6cm}p{0.8cm}p{0.8cm}p{0.6cm}}
object ID & RA, h,m,s & Dec, $^\circ\,\,^\prime\,\,\arcsec{\,}$ & D & $A_V$ & $R,\arcmin{\,}$ & \otoh & optical counterpart  &  \multicolumn{3}{c}{MPFS observations} & \multicolumn{3}{c}{SCORPIO observations} \\
&&&&&&&& $t_{exp}$ & sp. range  & res. &$t_{exp}$ & sp. range & res.\\
IC342 X-1   &   03 45 55.68     &   +68 04 54.9 &      3.28\footnote{Saha \etal (2002)}  &   1.51	& 14.86 &  8.85 (8.25) &  shock-powered nebula (Roberts \etal, 2003), $\sim 100pc$&  6900 & 4000-6900 & 5 & 7200 & 3900-5700 & 5 \\
%IC342 X-2   &    03 46 15.71      &	+68 11 12.2    &      3.28\footnote{\cite{Saha}}  &   1.85	&  ?        &  MPFS (2004/11/15?), SCORPIO (2005/11/09)\\
HoII~X-1   &    08 19 28.99  &	+70 42 19.4    &      3.39$^b$  &  0.105	& 13.74 &  7.92  &  HII region with strong HeII$\lambda$4686 emission (Lehmann \etal, 2005)   &  1800 & 4000-6900 & 5 & \multicolumn{3}{c}{---}\\
HoIX~X-1   &   09 57 53.28   &  +69 03 48.4    &      3.7\footnote{Karachentsev \etal (2002)} &  0.261  & 16.06 & [Fe/H]=-0.4$\div$-0.7\footnote{Tully (1988)} &  bubble nebula MH9/10, 200$\times$400~pc (Miller, 1995)     &  4500 & 4000-6900 & 5 & 1200 & 3460-7460 & 10 \\
% NGC4559 X-7 &	12 35 51.72  &   +27 56 04.7   &      10.3  \footnotemark[3] & 0.058 & 22.49 & 8.48;6.92  &  inside an annular star-forming region \cite{soria4559}  &  5400 & 4000-6900 & 5 & 5400  &  3900-5700 & 5\\
NGC5204 X-1 &    13 29 38.62  &  +58 25 05.6   &      4.5\footnote{Tully \etal (1992)}  & 0.04 & 1.96 & --- & several blue stars and nebular emission (Liu \etal, 2004)  &   8400 & 4000-6900 & 5 & 2700  &  3900-5700 & 5\\
M51 X-7 &     13 30 00.99  &  +47 13 43.9   &     8.4\footnote{Feldmeier \etal (1997)}  & 0.117 & 30.85 &  8.92 (6.72) & cluster with offset nebular emission  &   3600 & 4000-6900 & 5 & \multicolumn{3}{c}{---}\\
M101 P098 &     14 03 32.40    &  +54 21 03.1   &     7.2\footnote{Stetson \etal (1998)} & 0.029  & 2.54 &  8.8 (8.64)  & high-excitation nebula  &  4500 & 4000-6900 & 5 & \multicolumn{3}{c}{---} \\
NGC6946~X-1 &     20 35 00.75  &  +60 11 30.9  &     5.5\footnote{Hughes \etal (1998)}  & 1.141 & 33.31 & 8.7 (7.09) & peculiar nebula MF16 (Blair \etal, 2001)  &    5830 & 4000-6900 & 5 & 5400  &  3900-5700 & 5 \\
NGC7331~X-1 &  22 37 06.75  &   +34 26 17.6  &     15.1\footnote{Makarova \etal (2002)} &  0.301 &  19.79 &  8.68 (7.02) & young star cluster + HII region P98 (Petit, 1998)  & \multicolumn{3}{c}{---} & 2700 & 3460-7460 & 10\\
\end{tabular}
%\footnotetext[1]{Saha \etal (2002)}
%\footnotetext[2]{Karachentsev \etal (2002)}
%\footnotetext[3]{Tully (1988)}
%\footnotetext[4]{Tully \etal (1992)}
%\footnotetext[5]{Feldmeier \etal (1997)}
%\footnotetext[6]{Stetson \etal (1998)}
%\footnotetext[7]{Hughes \etal (1998)}
%\footnotetext[8]{Makarova \etal (2002)}
}
\normalsize

\end{sidewaystable*}

%%%All the spectra have moderate resolution $R \sim 1000$. 
%%%roughly constant in terms of $\Delta \lambda$ throughout the spectral range. 
MPFS was used with the grating \#~4 (600~lines/mm), 
%%%, having spectral resolution   $R \simeq 2000$ and spectral range $3900-7000\AAA$. 
SCORPIO was used with grisms VPHG550G and VPHG1200G. Spectral ranges and spectral resolutions 
are listed in table~\ref{tab:ulxlist}.   
%%The spectrograph transparency drops dramatically beyond $\lambda \simeq 3900\AAA$ resulting 
%%in additional spectral range limitation for VPHG550G. 
Seeing conditions during observations were from $1$ to $2\arcsec{\,}$. 
For the long-slit observations we estimate the light losses on the slit, consistent 
with the actual seeing value.
For the sources listed in table~\ref{tab:ulxlist} we have obtained either long-slit or 
panoramic spectra appropriate for analysis. 

Data reduction system was written in IDL6.0 environment, using procedures written by V. Afanasiev, A. Moiseev and P. Abolmasov. 
For long-slit observations the reduction system contains all the standard procedures.
For 3D data the standard procedures for panoramic data reduction were used (i. e. those standard for long-slit data plus extraction of fiber spectra and fiber sensitivity calibration). 
For MPFS data atmospheric dispersion correction was made with accuracy $\sim 0\arcsec{.}1$.

%%!!!!!!!!!!! opisat kratko obrabotku dannyh!!!!

In all the targets emission lines from the surrounding nebulae were detected. 
%%the cases integral spectra where obtained containing emission lines, in some cases 
Contribution from the stellar continua (host cluster or association) has also been detected in some cases. 
Emission line parameters were calculated using gauss analysis.  Fitting by
two gaussians was used for [SII]$\lambda\lambda$6717,6731 doublet,
H$\gamma$+[OIII]$\lambda$4363, HeII $\lambda$4686+FeIII$\lambda$4658  and 
FeII+FeIII $\lambda$5262,5270 blends, triple gaussian was used to deblend H$\alpha$ with
[NII]$\lambda$6548,6583 doublet (for the components of the
latter both fixed wavelength difference and flux ratio 
$F(\lambda 6583 / \lambda 6548)=3$ were adopted). Exact wavelength values 
with an accuracy of $\sim 0.1\AA$ were taken from Coluzzi (1996).
%Emission line integral fluxes are given in table~\ref{tab:lines}.

\begin{sidewaystable*}\label{tab:lines}
\caption{Observed H$\beta$ fluxes, emission-line fluxes in H$\beta$ units and unreddened luminosities of the ULXNe.
For NGC6946~X-1, HoII~X-1, M101~P098 and NGC5204~X-1 the luminosities were unreddened using
H$\alpha$/H$\beta$ criterion and Cardelli \etal (1998) reddening curves. For NGC7331~X-1 the best-fit interstellar absorption 
($A_V = 1\mag{.}4$) was derived from the stellar population spectrum fit. In other 
ULXNe the galactic $A_V$ was used (compare with table~\ref{tab:ulxlist}. 
%%Also the actual value of $A_V$ used for unreddening is given (cf. galactic $A_V$ in table~\ref{tab:ulxlist}).
In NGC7331~X-1 the luminosities in residual emissions are given in brackets (see text for details).
The total power estimates for HoII~X-1, NGC6946~X-1, M101~P098 and NGC5204~X-1 are 
hydrogen-ionizing luminosities derived in assumption that H$\beta$ is a recombination line 
(see section~\ref{sec:mf16}). 
Remaining nebulae are treated as shock-powered, their total powers were estimated 
using equation (\ref{E:ltot}).
%Emission line fluxes and luminosities of selected ULX counterparts. 
%Unreddened fluxes and line ratios are given.
%For MF16, HoII~X-1, M101~P098 and NGC5204~X-1 luminosities were unreddened using 
%H$\alpha$/H$\beta$ criterion. 
%For NGC7331~X-1 best-fit interstellar absorption ($A_V = 1\mag{.}4$) obtained 
%for the stellar population spectrum was used.
%For the other sources Galactic $A_V$ was used.
%For NGC7331~X-1 luminosities residual emission is given in brackets.
%Total power for HoII~X-1, NGC6946~X-1, M101~P098 and NGC5204~X-1 residual emission is 
%the hydrogen-ionizing luminosity in assumption that H$\beta$ is a recombination line 
%(see section~\ref{sec:mf16}). 
%Remaining nebulae are treated as shock-powered, and total power is estimated using equation~\ref{E:ltot}.
}
\footnotesize
\begin{tabular}{lcccccccc}
 \,                    & HoII~X-1  &  NGC6946~X-1  &  HoIX~X-1    & IC342~X-1 & NGC7331~X-1  &  M101~P098  & NGC5204~X-1 &  M51~X-7\\
F(H$\beta$), 
  $10^{-15}$\ergf      & 11        & 4.66$\pm$0.13 & 12.6$\pm$0.6 & 4.3$\pm$0.08     &  1.52$\pm$0.07  &  0.5$\pm$0.2   & 1.0$\pm$0.3 & 0.67$\pm$0.13  \\
   $A_V$               & 0.19      &  1.34         &  0.26      &     1.51        & 1.40     &  0.41   &  1.22 & 0.12 \\       
\hline   
H$\delta$    & --        & 0.16$\pm$0.13 &  0.40$\pm$0.15       &    --            &0.05$\pm$0.07 & ---  &  --- &   --- \\
H$\gamma$    & 0.45      &0.381$\pm$0.015&  0.31$\pm$0.10       &   0.28$\pm$0.02    &0.34$\pm$0.02 & --- & ---  & --- \\
\oiii $\lambda$4363 
                       & 0.08      &0.174$\pm$0.019&  0.133$\pm$0.15&    ---  &0.03$\pm$0.01 &  --- &  --- & ---\\
\hei $\lambda$4471 
                       & 0.08      & 0.06$\pm$0.06 &  --    &   0.051$\pm$0.015  &   ---       &  3$\pm$1 &  --- & 0.6$\pm$0.5\\
\heii $\lambda$4686
                       & 0.14      &0.170$\pm$0.015&  0.06$\pm$0.02&   0.036$\pm$0.015    &   ---     &  --- & --- & --- \\
\hei $\lambda$4713 
                      & 0.01      &      ---      &   --   &     --           &     ---   &  --- &  --- &  --- \\
\hei $\lambda$4922 
                     & 0.01      &      ---      &   --   &     --           &     ---   & ---  & --- & --- \\
\oiii $\lambda$4959
                     & 1.00      & 2.25$\pm$0.05 &   0.537$\pm$0.06 &   0.219$\pm$0.015  &0.35$\pm$0.04  &  1.7$\pm$0.3 & 2.0$\pm$0.2 & 0.47$\pm$0.14\\
\oiii $\lambda$5007
                     & 3.00      & 7.06$\pm$0.06 &  1.47$\pm$0.04 &  0.837$\pm$0.016   &1.15$\pm$0.05  &  5.6$\pm$0.3 & 5.8$\pm$0.3  & 0.92$\pm$0.16\\
\ni $\lambda$5200
                     & ---       & 0.15$\pm$0.05 &  0.07$\pm$0.03 &     0.302$\pm$0.014   & 0.08$\pm$0.04  &  0.3$\pm$0.1 &  --- & ---\\
\heii $\lambda$5412
                     & 0.02      & 0.14$\pm$0.05 &  ---  &      ---       &     ---      & --- & --- & ---\\
\nii $\lambda$5755
                     & ---       & 0.118$\pm$0.15&  0.04$\pm$0.02&     --    & 0.02$\pm$0.03  &  --- & --- &  ---\\
\hei $\lambda$5876
                     & 0.07      & 0.11$\pm$0.02 &  0.17$\pm$0.04 &     2.6$\pm$0.4    & 0.13$\pm$0.04 & --- &  --- & ---\\
(\oi $\lambda$6300+
\siii $\lambda$6310)
                     & 0.11      & 1.42$\pm$0.08 &   0.92$\pm$0.14  &    2.3$\pm$0.4    & 0.15$\pm$0.04 & --- & --- & ---\\
\oi $\lambda$6364
                     & 0.03      & 0.48$\pm$0.03 &  0.34$\pm$0.06  &     0.50$\pm$0.15   & 0.02$\pm$0.02 & --- & --- & ---\\
H$\alpha$            & 3.20      &4.728$\pm$0.087&  3.79$\pm$0.10  &    5.8$\pm$0.1       &6.375$\pm$0.009 & 3.45$\pm$0.10 & 4.5$\pm$0.5&  5.18$\pm$0.10\\
\nii $\lambda$6583
                     & 0.08      & 4.16$\pm$0.08 &  1.18$\pm$0.08    & 4.5$\pm$0.1   &2.00$\pm$0.01  &  1.24$\pm$0.10 & 0.6$\pm$0.1 & 2.9$\pm$0.2\\
\hei $\lambda$6678
                     & 0.03      & 0.06$\pm$0.03 &   $\sim$1?    &  0.2$\pm$0.1       &0.04$\pm$0.03  &  --- & --- & ---\\
\sii $\lambda$6717
                     & 0.32      & 2.46$\pm$0.02 &  1.73$\pm$0.04  &   3.6$\pm$0.2    & 0.97$\pm$0.02 &  0.6$\pm$0.2 & 1.6$\pm$0.3&  1.29$\pm$0.15\\
\sii $\lambda$6731
                     & 0.29      & 2.35$\pm$0.02 &  1.2$\pm$0.04  &    2.6$\pm$0.2  & 0.71$\pm$0.02 & 0.6$\pm$0.3 &  1.1$\pm$0.3 & 0.97$\pm$0.18\\
\hline
%\multicolumn{6}{c}{Luminosities, $10^{37}$\ergl:}\\
Luminosities, $10^{37}$\ergl:\\
% H$\delta$       & ---     & 1.42$\pm$0.07   & 1.2$\pm$0.5    &                &  4.5$\pm$0.3                                                                                    (0.7)\\
%H$\gamma$        & 0.75      & 3.28$\pm$0.12 &  0.88$\pm$0.02     &                &  9.5$\pm$0.5                                                                                    (2.7)\\
% \oiii$\lambda$4363& 0.13      & 1.25$\pm$0.15 &  0.38$\pm$0.02    &                &  0.9$\pm$0.6                                                                                    (0.9)\\
HeII$\lambda$4686 & 0.22      & 1.31$\pm$0.10 &    0.07$\pm$0.02  &    0.1$\pm$0.02     &  $\lesssim 1$   & $\lesssim 0.2$ &  $\lesssim 0.1$\\
H$\beta$           & 1.67      & 7.2$\pm$0.3   &   2.73$\pm$0.13  &      3.3$\pm$0.6     &  21.45$\pm$0.19 (6.9)  & 0.3$\pm$0.1 & 0.9$\pm$0.1 &   0.6$\pm$0.1\\
%\oiii$\lambda$4959,5007& 1.67      & 15.64$\pm$0.10 & 1.46$\pm$0.16       &                &  6.28$\pm$0.17                                                                                    (4.0)\\
\oiii$\lambda$5007  & 5.01      & 48$\pm$3 &  3.97$\pm$0.12$\pm$      &   1.4$\pm$0.3       &  19.72$\pm$0.19 (12.7) & 1.7$\pm$0.5 &  5$\pm$1 & 0.6$\pm$0.1\\           

\sii$\lambda$6717,6731  &   1.02   & 35$\pm$1  &  7.3$\pm$0.4    &    10.5$\pm$0.5   &  18.2$\pm$0.5 (17.2) &   0.4$\pm$0.2 & 1.6$\pm$0.5 & 1.4$\pm$0.2\\      
Total power             &   70     &  500      &  300            &    300           & \,\,\,\,\,\, (700) &    12  & 60 & 70 \\
\end{tabular}
\normalsize

\end{sidewaystable*}

%%% perenes sverhu
%%%Among these 
Among our targets there are two large-scale bubble nebulae (HoIX~X-1 and IC342~X-1) and four 
high-excitation nebulae (HoII~X-1, NGC6946~ULX-1, M101~P098 and NGC5204~X-1)
In two cases (M51~X-7 and NGC7331~X-1) comparatively high S/N cluster spectra with 
nebular contribution where obtained.

%%%%%%%%%%%%%%%%%%%%%%%%%%%%%%%%%%%%%%%%%%%%%%%%%%%%%%%%%%%%%%%%%%%%%%%%%%%%%
\section{Results and Interpretation}\label{sec:smod}

Emission line integral fluxes and luminosities are given in table~\ref{tab:lines}. 
In the last row we estimate the total power of the nebula suggesting it either a 
shock-powered (see section~\ref{sec:bubbles}) or photoionized HII region
(see section~\ref{sec:mf16}).

\subsection{Bubble nebulae}\label{sec:bubbles}

\begin{figure*}
\epsfig{file=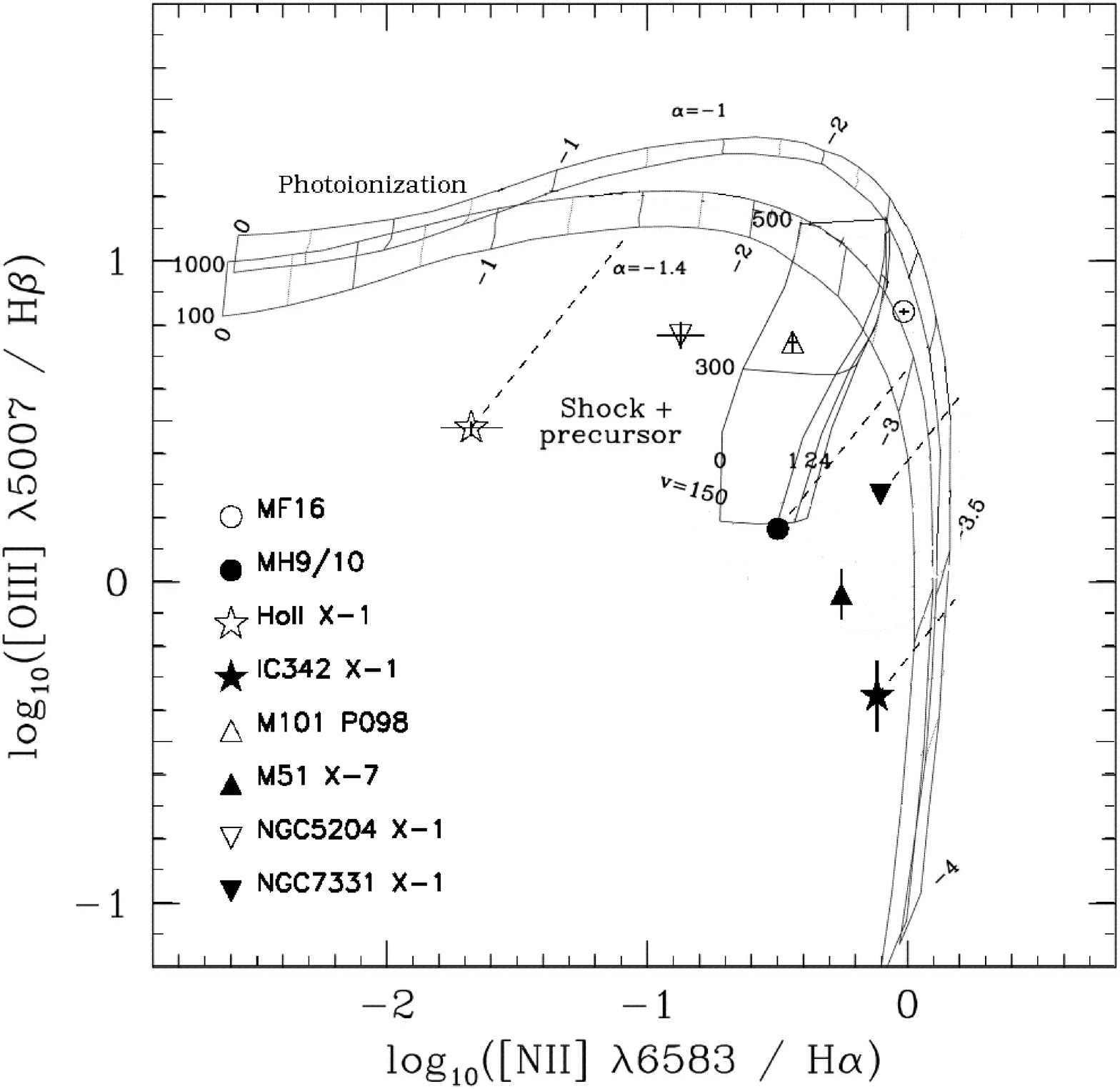,width=\textwidth,clip=1}
\caption{Observed ULXNe on ionization diagram [NII]6583/H$\alpha$ vs [OIII]$\lambda$5007/H$\beta$. 
Dotted lines indicate displacements from the observed position to that with solar abundance and  
the same physical conditions. 
Model grids are taken from the work of Evans \etal (1999). 
The longest curves stretching from the lower right to the upper left corner are MAPPINGsII  
photoionization models with a power-law source, labelled by logarithm of ionization parameter, 
ranging from -4 to 0, and by hydrogen density (100 and 1000 $cm^{-3}$, at the upper 
left corner). Shock+precursor MAPPINGsII models are shown labelled with the shock velocity in 
\kms and magnetic parameter (0, 1, 2, 4, for details see Evans \etal (1999) and Allen \etal (1998)).
}
\label{fig:evans}  
\end{figure*}

We classify HoIX X-1 MH9/10 nebula and IC342~X-1 nebula as purely shock-powered nebulae,
with the shock velocities in the range $20-100$\kms.
Both estimates follow from the kinematical data, however, in this paper we
concentrate on emission line intensities and diagnostics, leaving kinematical studies to 
further papers. If one suggests photoionisation for these two ULXNe, a source with 
$L \gtrsim 10^{42}$~\ergl\ is required.
Their positions on ionization diagrams (see Fig.~\ref{fig:evans}) are consistent with shocks of moderate power ($V_S \sim 100$\kms).
The total power of a nebula and the mean shock velocity (in the case when the physical size is known) can be estimated using its total H$\beta$ luminosity.
According to Dopita \& Sutherland~(1996), total energy flux in a shock wave:

\begin{equation}\label{E:total}
F_{tot} = 2.28 \times 10^{-3} V_2^3
n_0 \, erg \,cm^{-2} s^{-1}
\end{equation}

\noindent
where $V_2$ is the shock velocity in  $100$~\kms units, $n_0$ is the pre-shock hydrogen density in $cm^{-3}$.
For the energy flux in H$\beta$:

\begin{equation}\label{E:hbeta}
 \begin{array}{l} 
F_{H\beta} = F_{H\beta,shock}+F_{H\beta,precursor} = \\ 
\qquad{}
\left(  7.44 \times 10^{-6} V_2^{2.41} + \right. \\
% \\
\qquad{}
\left. +9.86 \times 10^{-6} \, V_2^{2.28} \right)
\,  n_0 \, erg\, cm^{-2} s^{-1},\\
 \end{array}
\end{equation}

\noindent
$F_{H\beta,shock}$ represents the emission of the cooling material behind the shock, $F_{H\beta,precursor}$ is the contribution of the precursor. 
The $F_{H\beta,shock}$ term is appropriate for any steady one-dimensional shock but the precursor term goes to zero at $V_S \lesssim 150$\kms.
Combining (\ref{E:total}) and (\ref{E:hbeta}) one expresses the H$\beta$ luminosity as a function of the total shock power and shock velocity:

\begin{equation}\label{E:ltot}
 \begin{array}{l} 
  L_{H\beta} = \left( 3.26\times 10^{-3} \,V_2^{-0.59} +\right. \\
\qquad{} \left. +4.32\times 10^{-3} \, V_2^{-0.72} \Theta(V_2-1.5) \right) L_{tot}\\
 \end{array}
\end{equation}

\noindent
where $\Theta(x)=1$ if $x \ge 0$ and $\Theta(x)=0$ otherwise.
From this equation the total power can be roughly estimated as $L_{tot} \sim (100 \div 300) L_{H\beta}$. 

\begin{figure*}
\epsfig{file=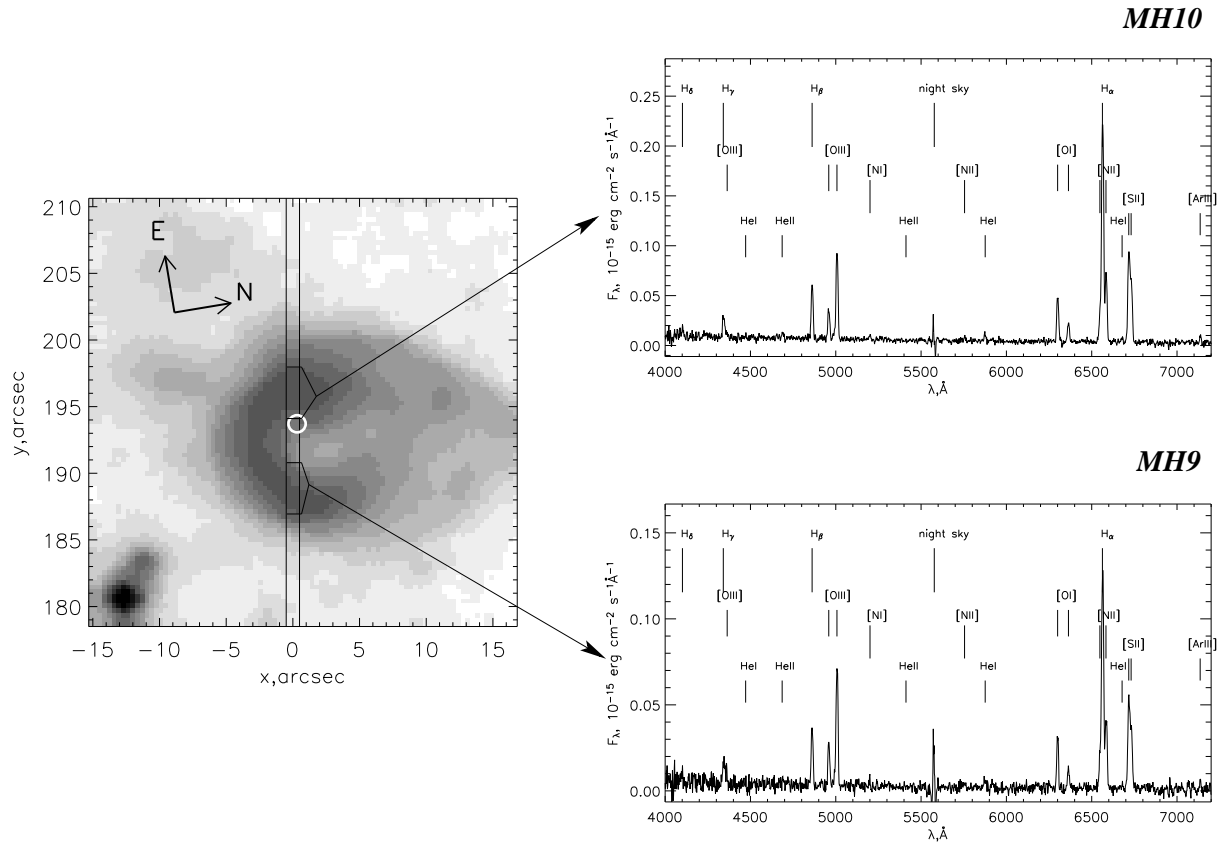,width=\textwidth,clip=1}
\caption{MH9/10 (the nebula associated with HoIX~X-1)  H$\alpha$ image (taken  
with SCORPIO) with the 1\arcsec slit. The X-ray source coordinates 
are marked by a white circle. To the right the spectra of the two parts of the nebula known 
as MH9 (bottom-right panel) and MH10 (top-right panel). In the upper spectrum \heii\,$\lambda$4686 
emission was detected.}
\label{fig:hoixfield}  
\end{figure*}

\begin{figure}
\epsfig{file=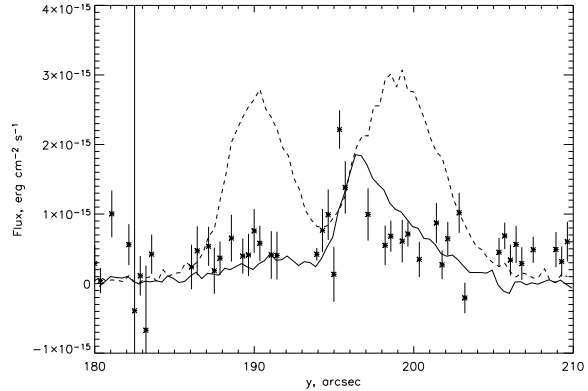,width=\columnwidth,clip=1}
\caption{1D maps for MH9/10 nebula (HoIX X-1) derived along the slit in 
continuum (solid line; integrated over the wavelength range $4000-7000\AAA$, 
emission lines excluded by median smoothing), H$\alpha$ (dashed line) and \heii $\lambda$4686 
(asterisks with error bars).}
\label{fig:hoixhe2}  
\end{figure}

In SCORPIO spectra of MH9/10 we detect roughly a point-like \heii $\lambda$4686 emission 
(figure~\ref{fig:hoixfield}) close to the
%coincident with the 
blue knot (central star cluster of the bubble) and the X-ray source (Gris\'e \etal, 2006). 
In figure~\ref{fig:hoixhe2} we present 1D maps of the X-ray source vicinity demonstrating that the
\heii\ line is emitted at the edge of the central cluster. 
We estimate the luminosity of the \heii $\lambda$4686 source as $\sim (2\div 4) 10^{36}$\ergl 
(the obtained value was multiplied by a factor $2-4$ estimating the losses of the slit). 
This is consistent with the results obtained by Gris\'e \etal~(2006), who also detect the \heii\ emission.

In HoIX~X-1 nebula this line may have stellar origin.
Possible sources of bright \heii$\lambda$4686 lines are Wolf-Rayet stars (Conti \etal, 1983) and 
accretion disks. For WN stars the equivalent width can reach $400\AAA$.
V-band luminosity is about $10^{37} \div 10^{38}$\ergl for bright WN stars, 
so one WN star can be enough to explain the \heii$\lambda$4686emission. 

\begin{figure*}
\epsfig{file=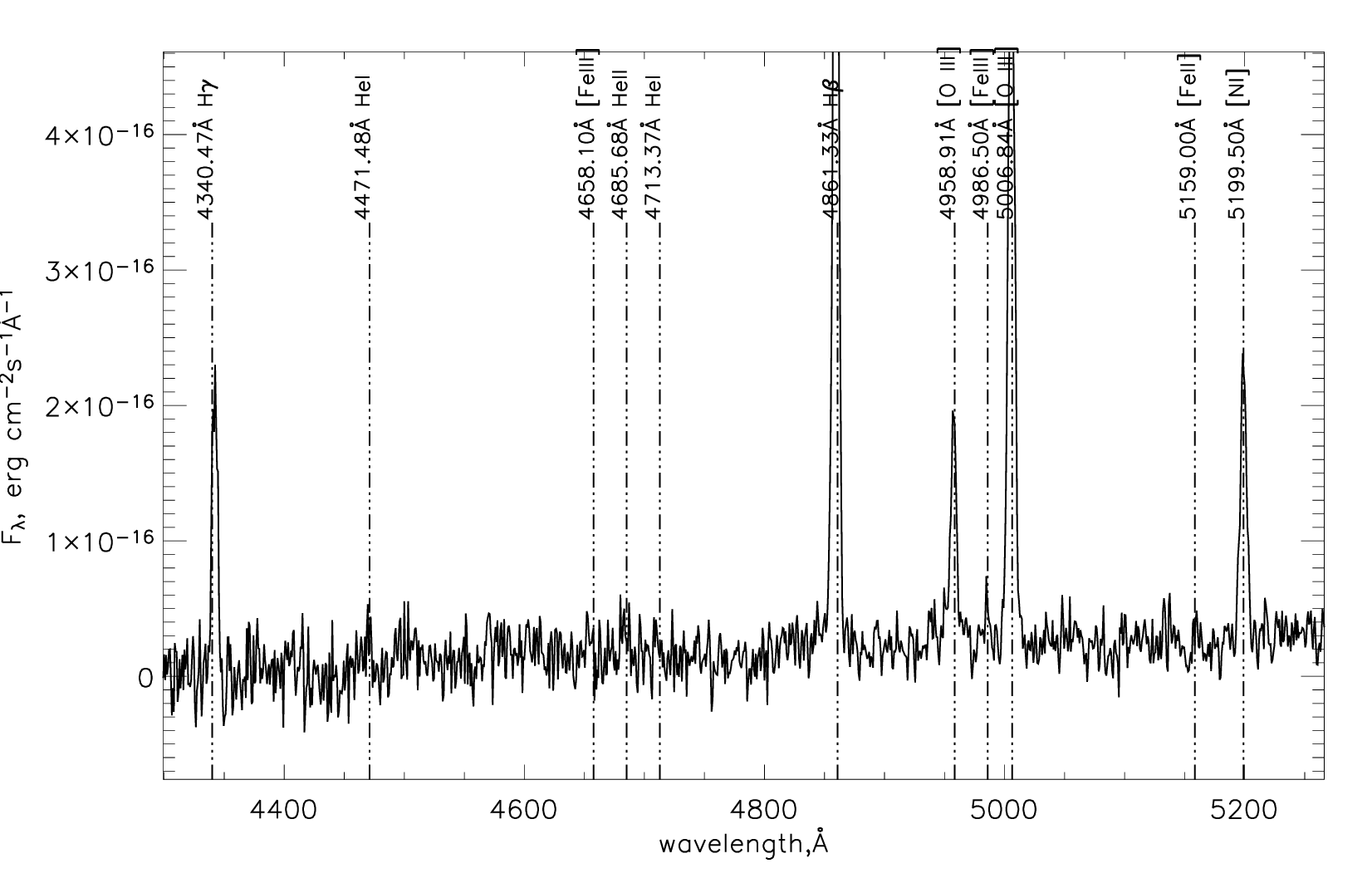,width=11cm,clip=1,angle=90}
\caption{Spectrum of the IC342~X-1 nebula (SCORPIO). The \heii $\,\lambda$4686 
emission originates roughly in the center of the ULXN. 
%High excitation and hydrogen lines are marked.
}
\label{fig:ic342sp}  
\end{figure*}

%% !!!!! risunok i text po IC342 X-1, kak min komentarii risunka + zhelatelno vyvod po superbubble
Bright emission lines in the IC342~X-1 nebula spectrum argue for shock excitation.
However, its integral spectrum (figure \ref{fig:ic342sp}) containes relatively faint 
high-excitation lines like \heii\ and \feiii\ emitted in the inner parts of the nebula. 
High-excitation spectrum resembles that of MF16 (see next section).

\subsection{NGC6946 X-1}\label{sec:mf16}

NGC6946~X-1 is known to coincide with a peculiar nebula MF16, for a long time considered a luminous Supernova Remnant (SNR).
% MF16 is peculiar even among the ULXNe.
It is the most compact (20pc$\times$34pc) and one of the brightest among the ULXNe.
Its integral spectrum is characterised by the highest \oiii$\lambda$5007/H$\beta$ 
and \nii$\lambda$6583/H$\alpha$ ratios. % (see figure~\ref{fig:evans}).

%Kinematical data do not show any particularly violent motions.
The velocity gradients and non-gaussian structure of the emission lines in our spectra suggest 
expansion velocity not greater than 200\kms. Dunne \etal (2000) have resolved H$\alpha$,
\nii\ and \oiii\ lines in a narrow (FWHM$\sim$20-40\kms) component from unshocked 
material and a broad (FWHM$\sim$250\kms) from shocked material. 
%, most likely about 100\kms. 
Basing on the MPFS and SCORPIO spectra we performed a thoroughful analysis of the 
ionization and excitation sources of the nebula (Abolmasov \etal, 2006; Abolmasov \etal, 2007a), leading to a 
conclusion that most of the power of the nebula comes not from shock waves but from 
photoionizing ultraviolet source even more luminous than the central X-ray source. 
CLOUDY modelling gives the parameters of the central source in the suggestion 
that the ionizing continuum in the 
EUV/UV is a black body: $T = 10^{5.15\pm0.15} K$, $L_{BB} = (7.5\pm 0.5) \times 10^{39}$\ergl. 
For the EUV luminosity one can make rough, but universal estimates (the lower limits), 
based on the ionizing quanta numbers. 
If \heii$\lambda$4686 and H$\beta$ are purely recombination lines one can apply the 
relations by Osterbrock~(1974):

\begin{equation}\label{E:he2ion}
\begin{array}{l}
L(\lambda < 228\AAA) 
\ge  \frac{4 \Ry}{E(\lambda\,4686)} 
\frac{\alpha_B(He^{++})}{\alpha^{eff}_{HeII\,\lambda4686}}
\times L_{HeII\lambda\,4686} \simeq \\
\\
\qquad{}\qquad{} \simeq 100 \, L_{HeII\lambda\,4686} \\
\end{array}
\end{equation}

\begin{equation}\label{E:hion}
L(\lambda < 912\AAA) \ge  \frac{1 \Ry}{E(H\beta)} 
\frac{\alpha_B(H^+)}{\alpha^{eff}_{H\beta}}
\times L_{H\beta}
\simeq 65 \, L_{H\beta},
\end{equation}

\noindent
%where !!!! ob'yasnit alphas!!
Here $\alpha_B$ values are total recombination 
rates for the Case B (optically thick to ionizing quanta)
for hydrogen and ionized helium, correspondingly. $\alpha^{eff}(line)$ is the effective 
recombination rate with the particluar emission line quanta production. $\alpha_B(He^{++}) 
/ \alpha^{eff}_{HeII\,\lambda4686}$ and $\alpha_B(H^+) / \alpha^{eff}_{H\beta}$
ratios depend weakly on the plasma parameters. The former changes by about factor of two 
in the temperature range $( 0.3\div 10) \times 10^4 K$, the latter by about 20\%.

In table~\ref{tab:lines} we give the hydrogen-ionizing luminosities defined this way as estimates for the power of photoionized nebulae.
H$\beta$ line can overestimate the photoionizing flux because contribution from shocks may be significant.
Discussion about the nature of the EUV photoionizing source follows in section~\ref{sec:discuv}.

\begin{figure*}
\epsfig{file=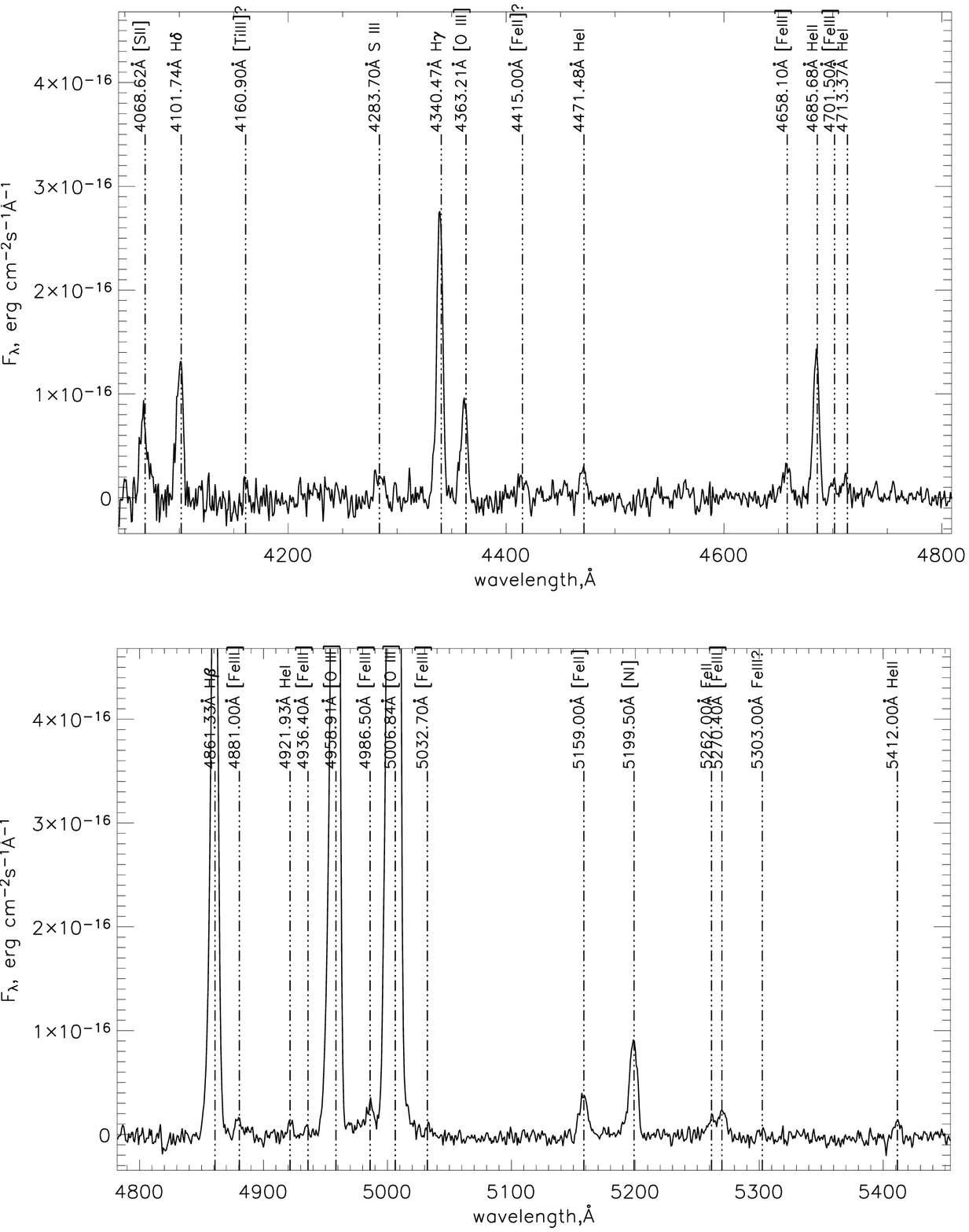,width=\textwidth,clip=1,angle=0}
\caption{
Integral spectrum of MF16 (NGC6946~ULX-1 nebula) taken with SCORPIO.
}
\label{fig:mf16sp}  
\end{figure*}

High signal-to-noise spectrum of MF16 obtained with SCORPIO shows numerous high-excitation lines like \feiii\ and Ar~IV. 
Integral spectrum is shown in figure~\ref{fig:mf16sp}.

\subsection{High-Excitation ULXNe}

\begin{figure*}
\epsfig{file=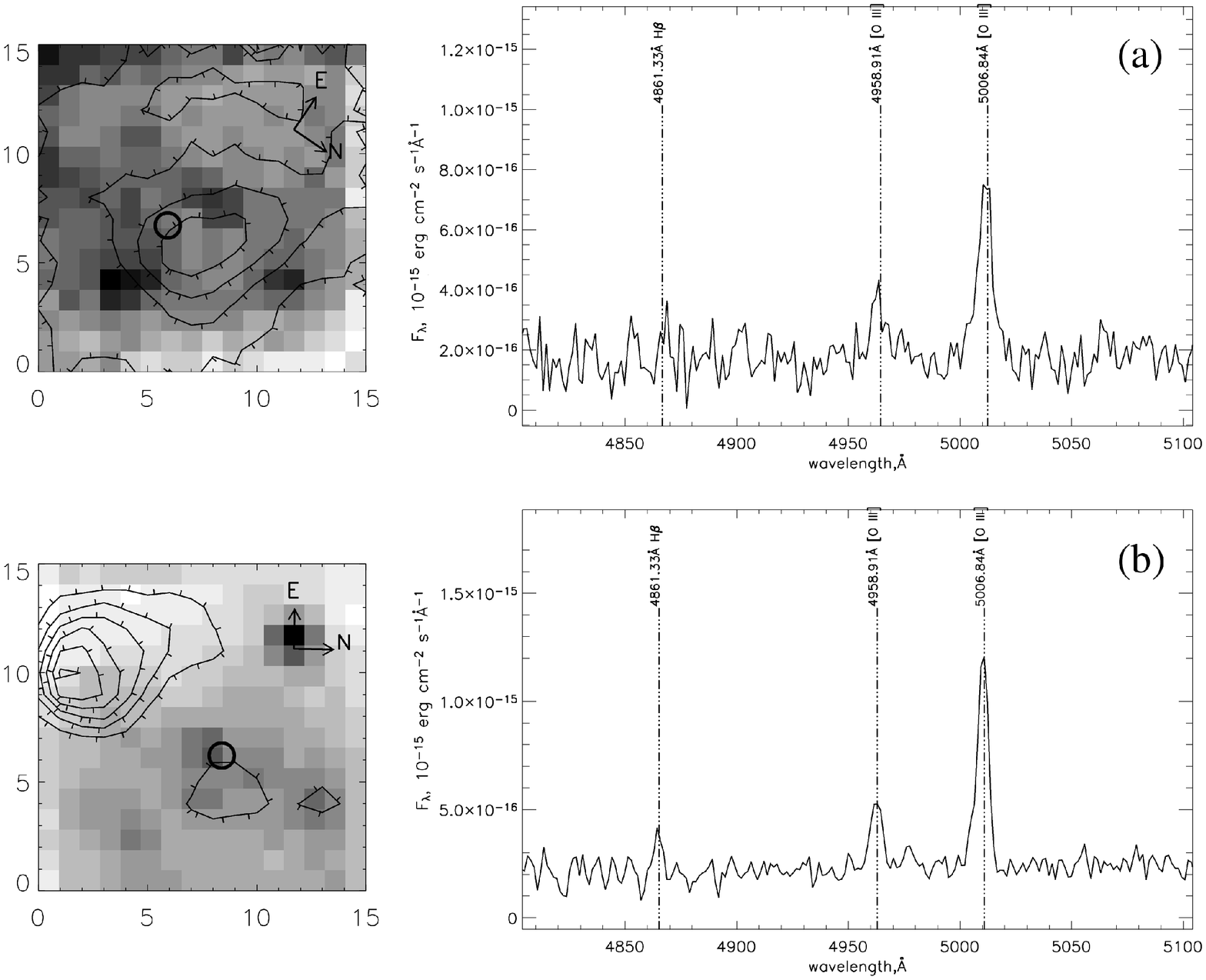,width=\textwidth,clip=1}
\caption{High-excitation ULXNe in M101 (a) and NGC5204 (b) in $16\arcsec{\,} \times 16\arcsec{\,}$ MPFS field. 
To the left continual flux maps 
in MPFS fields are shown overplotted with the \oiii$\lambda$4959,5007 flux contours. Here and below tick marks point towards lower intensity regions.
To the right integral spectra of the \oiii-bright regions adjacent to the X-ray sources are shown.
Chandra X-ray positions are shown by circles.}
\label{fig:o3bright}  
\end{figure*}

In the vicinity of two of the observed sources (M101~P098 and NGC5204~X-1) we detect 
regions with high \oiii$\lambda$5007/H$\beta$ ratio ($>5$). 
They can be seen as bright spots on the \oiii$\lambda$5007 flux maps shown in 
figure~\ref{fig:o3bright}. % together with the integral spectra in the H$\beta$ -- \oiii$\lambda$5007 region. 
The \oiii$\lambda$5007/H$\beta$ ratios are similar to that of MF16, but the integral luminosities 
are much less -- an order of magnitude less for NGC5204~X-1 and even more for M101~P098.

M101~P098 nebula is well resolved as an elongated region $1\arcsec{\,}\times 4\arcsec{\,}$, 
bright in \oiii$\lambda$5007,4959 lines (see figure~\ref{fig:o3bright}a). 
For the distance 7.2 Mpc this corresponds to the physical size about $35pc \times 140 pc$. 

Bright  \oiii emission lines argue for either a photoionizing source capable to provide 
ionization parameter\footnote{We define ionization parameter as (for example, in Evans \etal (1999)): 
$$ U = \frac{1}{cn} \int^{+\infty}_{13.6eV}\frac{F_E}{E} dE,$$ 
where $n$ is the gas number density.}
 $\gtrsim 10^{-3}$ at about 50~pc or shock waves with $V_S \gtrsim 300$\kms propagating 
at similar spatial scales (Evans, 1999; Dopita \& Sutherland, 1996). 
Photoionizing source must have the hydrogen-ionizing luminosity $\gtrsim 2 \times 10^{38}$\ergl, 
so M101~P098 can be considered an analogue of MF16 in a more rarefied ISM. 
From the other hand, fast shocks with $V_S \sim 300-500$\kms require practically 
identical power $(1.5-2) \times 10^{38}$\ergl\ to provide the observed H$\beta$ luminosity. 
The example of MF16 suggests that these high-excitation nebulae are more likely to 
be photoionized HII regions. However, both photoionization and shock excitation can act in 
powering these nebulae.

\begin{figure*}
\epsfig{file=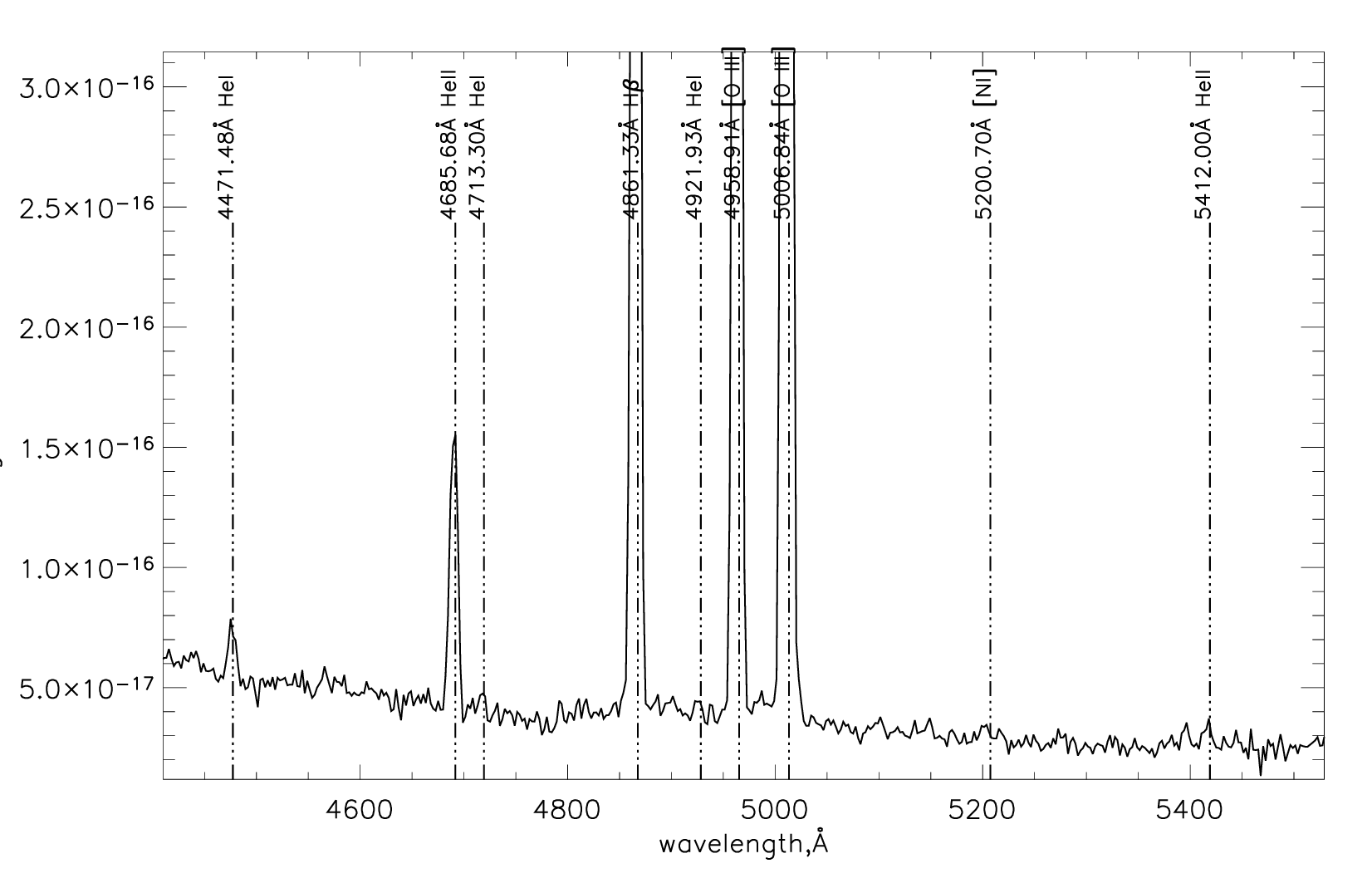,height=\textwidth,clip=1,angle=90}
\caption{
Integral spectrum of the HoII~X-1 nebula (Lehmann \etal, 2005). Only part of the spectrum in 
4500-5500$\AA$ spectral range is shown.
}
\label{fig:hoiisp}  
\end{figure*}

HoII~X-1 is another example of a high-excitation ULX nebula. 
\oiii$\lambda$5007/H$\beta \sim 3$ is low if compared with the  M101~P098 and NGC5204~X-1 spectra, 
but HoII has lower oxygen abundance, so the physical conditions may be quite similar.
\heii $\lambda$4686 emission line is very bright relative to H$\beta$ ($F(HeII\lambda4686) / F(H\beta) \sim 0.2$), 
but the total luminosity in this line is an order of magnitude lower than that in MF16. 
Our results on this object were published in Lehmann \etal (2005). 
Long-slit spectrum (obtained with UAGS spectrograph) of the nebula is shown in figure~\ref{fig:hoiisp}.

% !! Tut ris i kakie-to slova (hotya by po ris) pro NGC5204. Tozhe kak UV source.

\subsection{Star Clusters}

Among the observed ULXs two objects coincide with isolated bright star clusters: 
NGC7331~X-1 and M51~X-7. Among the other ULXs whose X-ray boxes contain stellar clusters, there are 
HoIX~X-1 (the cluster is very faint, $V \sim 20\mag{.}5$ and not massive) and 
NGC5204~X-1 (the cluster is a part of a more complex star-forming region).   
%%For HoIX~X-1 the cluster is very faint ($V \sim 20\mag{.}5$), and for NGC5204~X-1 it is a 
%%part of a more complicated star-forming region. 

\begin{figure*}
\epsfig{file=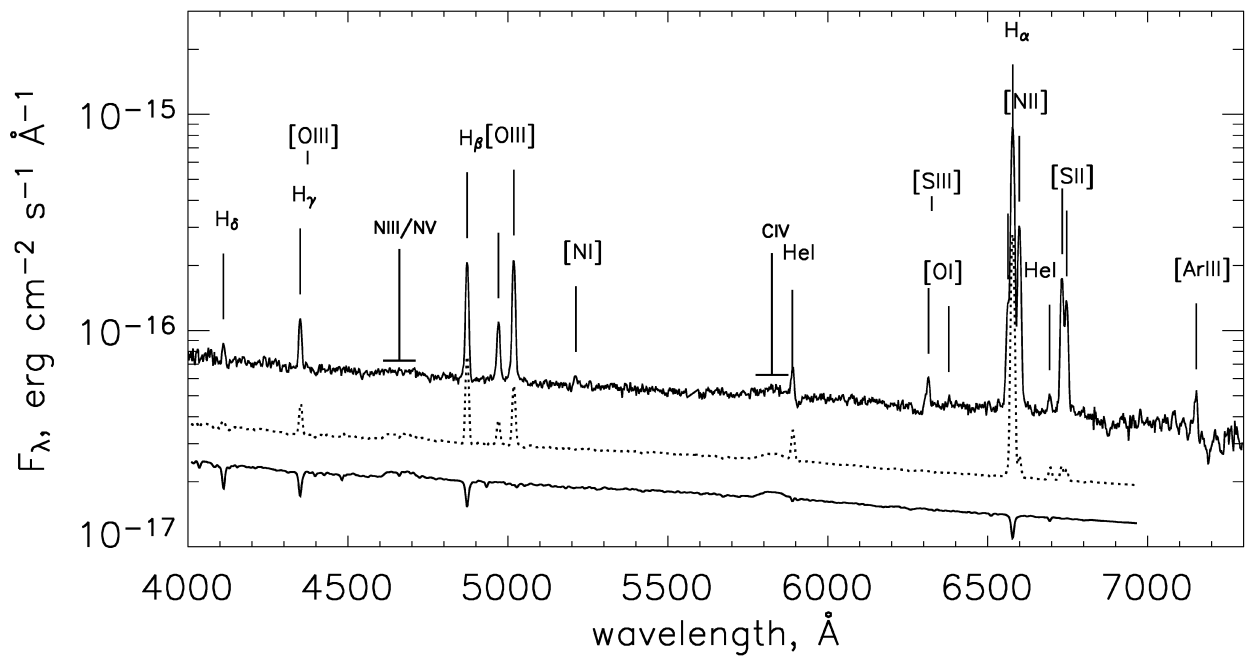,height=\textwidth,clip=1,angle=90}
\caption{Integral spectrum of the cluster in NGC7331~X-1 (upper solid), the best-fit StarBurst99 model 
(lower solid) and the best-fit StarBurst99 + best-fit nebular photoionization CLOUDY model spectrum. 
Model spectra are shifted in vertical direction.}
\label{fig:ngc7331}  
\end{figure*}

%NGC7331 is situated at about 15Mpc, and the 
A nearly point-like continual source is seen in ground-based images of the NGC7331~X-1 region.
At the distance of 15~Mpc its size is $\sim50$~pc. In the HST images the source consists of 
young cluster and some fainter sources around (Abolmasov \etal, 2007b).
% actually has size about 50pc and consists of several 
%clusters at some distance from the X-ray source. 
The cluster coincides with the HII region P98 (Petit, 1998).
Our study of the source environment based on {\it HST} and SCORPIO observations is
published in a separate paper (Abolmasov \etal, 2007b). 
We fitted the integral spectrum of the stellar population with StarBurst99 models (Leitherer \etal, 1999; V\'azquez \& Leitherer, 2005) 
with solar and 0.4 of solar metallicities, and derived the best-fit age $T=4.25\pm0.5~Myr$ and 
the cluster mass $M \sim 40 000 $\msun. 
The integral spectrum together with StarBurst99 best-fit stellar population and CLOUDY 
nebular spectra is shown in figure~\ref{fig:ngc7331}. 
WC features are bright in the spectrum, pointing to the presence of several WC stars. 
Emission line spectrum can not be totally explained by photoionization by the central cluster.
Low-excitation shock-ionized lines like \sii$\lambda$6717,6731 and 
\nii$\lambda$6548,6583 have strong excess that can not be explained by the contribution from stellar 
winds and supernovae.  % (for a cluster of such mass). 
We suggest that in this case the ULX is an additional source of 
mechanical power, comparable with the total luminosities of the brightest ULXNe like MF16.
The residual emission line luminosities are given in brackets in table~\ref{tab:lines}.

\begin{figure}
\epsfig{file=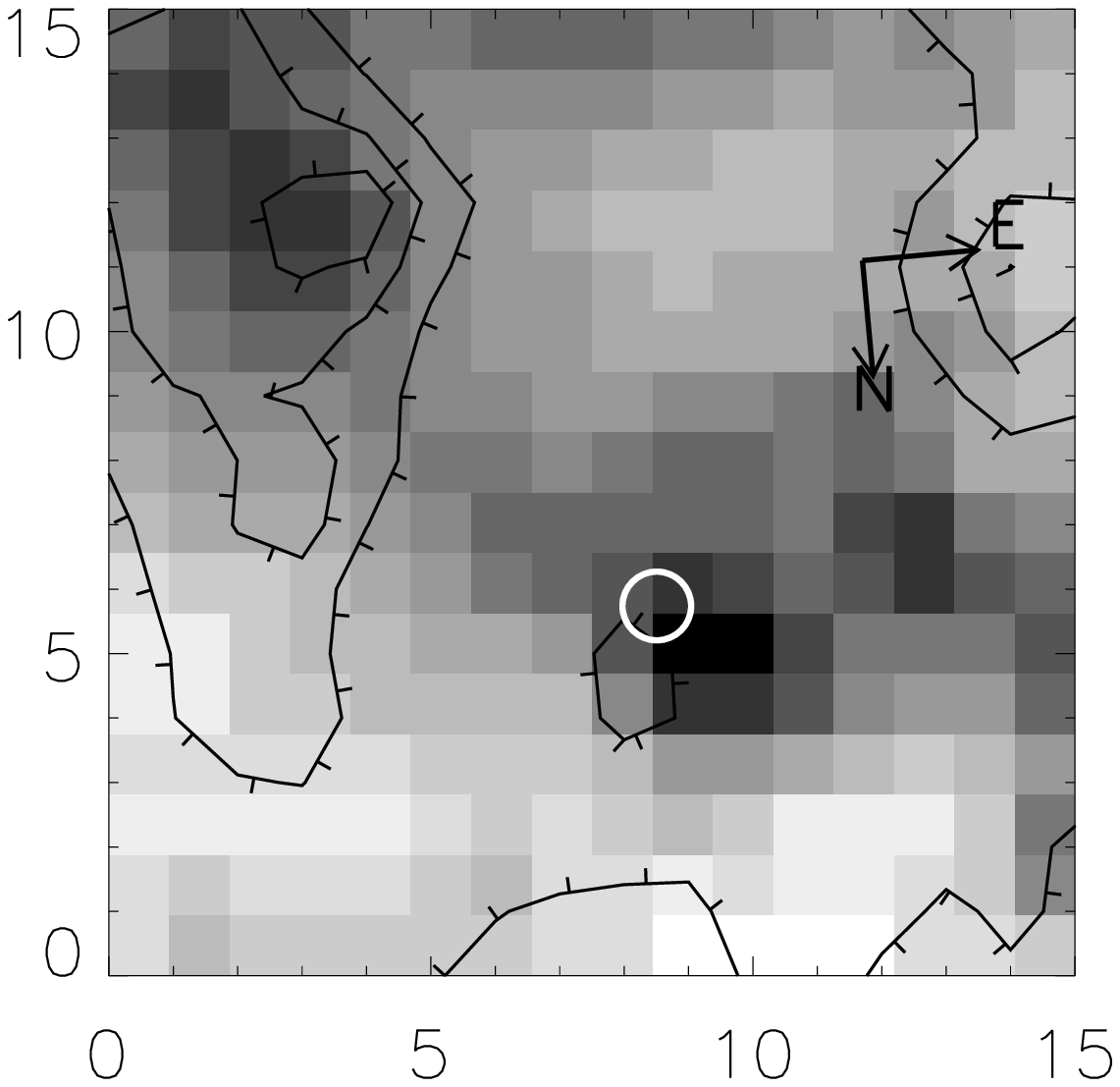,width=\columnwidth,clip=1}
\caption{M51~X-7 $16\arcsec{\,} \times 16\arcsec{\,}$ field observed with the MPFS. 
Stellar continuum integrated in the total spectral
range is shown with the H$\alpha$ countours overplotted. {\it Chandra} ULX position 
is marked by a white circle.}
\label{fig:m51reg}  
\end{figure}

M51~X-7 is located at the outskirts of a bright star cluster (figure~\ref{fig:m51reg}), 
clearly identified with a young massive cluster {\it n5194-839} from the list of Larsen (2000). 
It is a Super Star Cluster (SSC), or a young massive cluster, with the integral 
$M_V = -11\mag{.}09$ and $U-B = -0\mag{.}81$,
indicating a rather young object of $T \sim 12 \, Myr$.
However, our spectral fitting with StarBurst99 models gives much larger cluster age 
$T \sim 60\pm15 \, Myr$ and $A_V \sim 0\mag{.}5\pm 0.1$.
Most likely, stellar populations of different ages are seen in the vicinity of the X-ray source. 
H$\alpha$ emission accompanied with bright \nii$\lambda$6583,6548 and \sii$\lambda$6717,6731 
emission lines is coincident with the X-ray source within the spatial resolution. 
As emission line spectrum contains signatures of shock excitation we apply 
equation~\ref{E:ltot} for the estimate of the total power. A detail study of the M51 X-7 will be 
published in a separate paper.

\section{Discussion}\label{sec:trep}

\subsection{Metallicity and Oxygen Abundances}\label{sec:discmet}

In table~\ref{tab:ulxlist} we list oxygen abundance and metallicity estimates for 
the host galaxies and the ULX positions in the galaxies according to Pilyugin~\etal~(2003).
It was argued by Soria~(2006) that low metallicities can be crucial in understanding 
the nature of ULXs. 
Many of the objects are indeed found in sub-solar metallicity environment, most likely at 
about 20-50\% of solar metallicity.
However, in our study of the star forming region associated with NGC7331~X-1 we 
find that solar-metallicity models provide better fit to the cluster spectrum that subsolar 
metallicity models with $Z = 0.4 Z_\odot$ and less. 

In HII regions it is easier to estimate oxygen and nitrogen abundances rather than $[Fe/H]$. 
We can estimate oxygen abundance for some of our objects using the method of Pagel~(1992),
based on the similarity between oxygen and hydrogen ionization potentials.
The method is based on the three-level atomic solutions by McCall~(1984), allowing reliable calibrations of the ion abundances from their forbidden lines relative strengths. Oxygen abundance estimates use two relations for two ionization states of the element:

$$
\begin{array}{l}
12+\lg \frac{O^{+}}{H^{+}} = \lg \frac{I_{[OII] \lambda 3726+\lambda 3729}}{I_{H\beta}} +\\
\qquad{} +6.174 + \frac{1.251}{t_3} - 0.55 \lg t_3 \\
\end{array}
$$

$$
\begin{array}{l}
12+\lg \frac{O^{++}}{H^{+}} = \lg \frac{I_{[OIII] \lambda 4959+\lambda 5007}}{I_{H\beta}} +\\
\qquad{} 5.890 + \frac{1.676}{t_2} - 0.40 \lg t_2 +\\
+ \lg \left( 1 +1.35 \times 10^{-4} n_e t_2^{-1/2}\right),\\
\end{array}
$$

\noindent
where $n_e$ is the electron concentration in $cm^{-3}$.
$t_3$ and $t_2$ are the electron temperatures (in $10^4 K$ units) of the gas emitting in \oiii\
 and \oii\ lines, correspondingly. 
$t_3$ can be obtained from the characteristic line ratio 
$I_{[OIII] \lambda 4959+\lambda 5007} / I_{[OIII] \lambda 4363}$
sensitive to electron temperature (Osterbrock,~1974).
Because of the close ionization potential $t_2$ must be similar to the electron 
temperature estimate $t_n$ made from another characteristic line ratio, 
$I_{[NII] \lambda 6583+\lambda 6548} / I_{[NII] \lambda 5755}$. 
Usually, both \oii\ and \oiii\  lines are present in a single \hii\ region. 
Total abundance can be found by adding the ion abundances:

$$
12+\lg \frac{O}{H} \simeq 12+\lg \left( \frac{O^{+}}{H^{+}} + \frac{O^{++}}{H^{+}}\right) \\
$$

Unfortunately we do not detect \oii$\lambda$3727 line in our spectra because of the 
limited spectral range for MPFS and sensitivity decrease for SCORPIO in the near UV range, 
so some additional assumptions will be needed.

For MF16 we have temperature estimates $t_3 = 1.9 \pm 0.1$ and $t_2 \simeq t_n = 1.5 \pm 0.2$, and 
$n_e \simeq 500 cm^{-3}$ derived from the sulfur doublet \sii$\lambda 6717,6731$ components ratio. 
For MF16 oxygen abundance can not differ much from the solar, because both very bright
\oiii\ and \oi\ lines are well detected.
% there and their intensities give independently the same abundance of oxygen (\cite{ngc7331}).
For a very broad range of CLOUDY models (successfully predicting most of the observed line ratios) % \cite{mf16} have found 
$I_{[OII]\lambda 3726+\lambda 3729}/I_{[OIII] \lambda 4959+\lambda 5007} \sim 1$. 
Using this estimate together with the directly observed value $I_{[OIII] \lambda 4959+\lambda 5007} / I_{H\beta} \simeq 9$, the temperature and density estimates one finds for MF16  $12+\lg \frac{O}{H} \simeq 8.5$, i. e. practically solar abundance. 

% Basing on these ULXNe (in NGC7331 and NGC6946), where we estimated oxygen abundance, we may suggest that low elementary abundances are not mandatory for ULX formation. 

\subsection{UV Sources}\label{sec:discuv}

Point-like optical counterparts of ULXs are usually found as blue stars with 
$M_V \sim -5 \div -8\mag{\,}$ (Terashima \etal,~2006).
On the other hand, there are strong indications that at least some of the ultraluminous X-ray  sources 
must be also ultraluminous EUV sources (in the range $20 \AAA \lesssim \lambda \lesssim 1000 \AAA$ 
responsible for hydrogen, oxygen and helium ionization). 
In the framework of the two most popular models (IMBH binaries and supercritical accretors), we can expect two explanations for the 
bright EUV radiation:

\begin{itemize}
\item[-] Very massive ($M \gtrsim 10^4 $\msun) IMBH accreting at several percent of its
Eddington level. Less massive IMBHs have too hard standard disk spectra. Note that the donor star 
contribution in the EUV region is most likely negligible.
\item[-] Supercritical accretion disk (SCAD) with stellar mass black hole and
UV-radiating wind photosphere, like what we observe in SS433 (Fabrika,~2004). 
\end{itemize}

IMBHs in binary systems are expected to have nearly-standard disks (because the accretion 
rates are most likely in the range $0.01-1.0$) with low outer temperatures.
If the accretion disk in such system is tidally truncated by the secondary, its radius 
will be some fraction (roughly $0.5$ of the black hole Roche lobe radius, 
practically equal to the binary separation, for example see Blondin (2000)) of the binary 
separation, that is  $a \sim R_* \left( M_{BH} / M_* \right)^{1/3}$. 
This leads to at estimate of the outer disk temperature:

$$
T_{out} \simeq 10^{-3} T_{in} \left(\frac{R_*}{10 R_\odot} \right)^{-3/4}  
\left(\frac{M_{BH}}{10^3 M_\odot} \right)^{-1} \left(\frac{M_*}{10 M_\odot} \right)^{-1}
$$

\noindent
In UV and optical ranges IMBH accretion disks have approximately power-law spectra 
with $F_\nu \propto \nu^{1/3}$ (Shakura \& Sunyaev,~1973). 

According to Poutanen \etal \,(2006), the temperature of the outer photosphere of a supercritical 
disk wind can be expressed as functions of the black hole mass in solar units $m$ and 
mass accretion rate in Eddington units $\dot{m}$ as:

\begin{equation}\label{E:tph}
T_{ph} = \left\{ 
\begin{array}{cr}
0.2 keV~m^{-1/4} \dot{m}^{-1} &  for~v\propto r^{-1/2}\\
0.8 keV~m^{-1/4} \dot{m}^{-3/4} &  for~v=const,\\
\end{array}
\right.
\end{equation}

\noindent
where $v$ is the radial velocity of the outflowing wind forming inside of the spherization
radius, $v\propto r^{-1/2}$ 
corresponds to the case when the velocity is everywhere proportional to the virial, 
and $v=const$ is the case of a relatively fast wind. 
$m$ is the black hole mass in Solar units, $\dot{m}$ is the mass accretion rate in Eddington units. 

For accretion rates $\dot{m} \sim 10-100$ maximum of the photosphere radiation falls
in a spectral range of hundreds of angstroms. Expected total photosphere luminosity is $\sim 10^{39}\div 10^{40}$\ergl.

\begin{figure*}
\epsfig{file=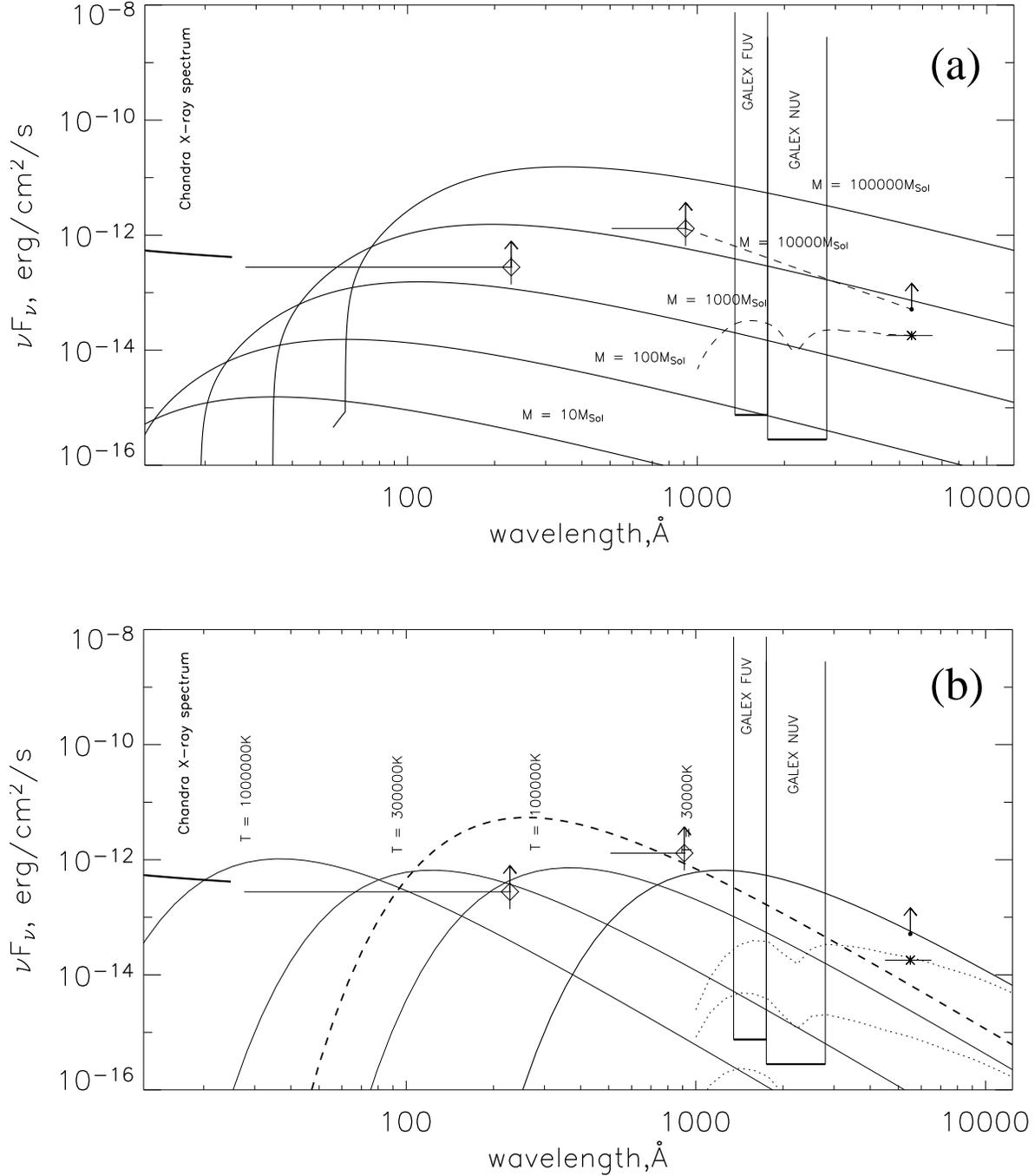,width=\textwidth,clip=1}
\caption{Spectral energy distribution from the X-rays to the optical range derived for
NGC6946~X-1. 
Optical source (star $d$, the V-band flux) is indicated by an asterisk, the upward arrow  
is the value corrected for the Galactic absorption (it is a lower limit because of possible 
intrinsic absorption).
Two diamonds with bars ranging towards higher energies are the estimates for the EUV luminosity 
based on the optical emission-line luminosities of MF16 
(HI and HeII lines, the symbols mark HI and HeII Lyman edges). 
The X-ray spectrum is represented by the best-fit model for the {\it Chandra} data 
given by Roberts \& Colbert~(2003). 
In two \textit{GALEX} bands the limiting fluxes
are shown with the $S/N=3$ detection limit for exposure time $t=10^5 s$. 
(a) MCD models for black holes accreting at $0.01 \dot{M}_{Edd}$ are shown with
solid curves. Dashed lines show both a power-law interpolation between 
hydrogen edge and V-band, and the same spectrum, but reddened by Galactic absorption
(Cardelli \etal (1998) reddening curves were used).
%H-ionizing and V-band luminosities are connected with a power-law shown by a thin dashed 
%line together with the corresponding spectrum reddened by Galactic absorption 
%(Cardelli \etal (1998) reddening curves are used).
(b) Black body models with $L=10^{39}$\ergl are presented, together with the optimal 
blackbody model for the UV radiation obtained in CLOUDY modelling. Dotted curves represent 
reddened black body spectra.
}
\label{fig:uvseds}  
\end{figure*}

In figure~\ref{fig:uvseds} we present spectral energy distributions (SEDs) for 
multicolor accretion disk (Shakura \& Sunyaev, 1973; Mitsuda \etal, 1984) models for IMBHs and SCAD photospheres with different black body temperatures. 
The NGC6946~ULX-1 SED inferred from optical and X-ray observations is also presented.  
The SED of NGC6946~ULX-1 looks fairly flat compared with blackbody and MCD models. 
This can be explained qualitatively by a supercritical accretor system spectrum 
(Fabrika \etal, 2007): the disk funnel ($\sim 10^{39}-10^{40}$~\ergl in X-ray band), 
the wind photosphere ($\sim 10^{39}-10^{40}$~\ergl in EUV/UV) and the donor star 
($\sim 10^{38}$~\ergl in the optical/UV).

If ULXs are indeed supercritical accretors one may expect a wide range of X-ray luminosities 
and photosphere temperatures, resulting in large variety of UV/EUV properties practically 
independent of the X-ray spectrum. 
For $\dot{m} \sim 100$ a bright EUV source and a photoionized nebula are likely to appear.
For $\dot{m} \lesssim 10$, however, one can expect only a compact HII region with 
relatively bright high-excitation lines like \heii $\lambda$4686 and \oiii$\lambda$5007. 
This is probably the case for HoII~X-1 and M101~P098.
Very large accretion rates ($\dot{m} > 1000$) result in a very soft photosphere emission 
combined with X-ray ionization and strong wind producing a large shock-powered nebula. 

As can be seen from figure~\ref{fig:uvseds} ULXs are reachable sources for {\it GALEX} (Martin \etal, 2003). 
For sources with lower absorption even UV spectroscopy will be possible in pointing observations. 
The main predicted difference between SCADs and IMBHs in the UV is in the 
difference of the spectral slopes in these two models.
We also expect  SCADs to be more divergent spectral properties in the UV range.

\section{Conclusions}

We confirm that ULXs belong to young stellar population. Some of them 
are certainly formed in star clusters and associations of 5-10\,Myrs old. Most of 
ULXs are surrounded with nebulae. Studying the nebulae gives us a powerful tool
to investigate the nature of ULXs. At least 4 (group A) of 8 ULXNe studied in this paper
show high flux ratio \oiii$\lambda5007$/H$\beta = 3\div 7$ (table~\ref{tab:lines}), that requires a strong EUV source
(ultraluminous UV sources, UUVs) to ionize the nebula. Other 4 of the 8 ULXNe (group B) 
are most likely shock-ionized.
%signatures of shock excitation, they have a ratio [SII]$\lambda6717,6731$/H$\beta = 2.3-4$. 

There is a ULXN in the group B, the nebula of NGC7331 X-1. 
Abolmasov \etal\ (2007b) have found parameters of the young super-stellar cluster (SSC) hosting the ULX
and calculated the nebula spectrum using CLOUDY photoionized by the SSC in NGC7331.
Residual emission line spectrum (values in brackets in table~\ref{tab:lines}), obtained after 
subtraction of the model spectrum from the observed spectrum, it shows both enhanced 
shock-excited lines (like \sii$\lambda$6717,6731) and \oiii$\lambda$4959,5007 doublet 
relative to the total emission-line spectrum of the cluster HII region. 
The residual spectrum of the nebula in NGC7331 is quite similar to spectrum of MF16, but 
about twice as fainter. NGC7331~X-1 residual emission may be a fainter equivalent of MF16.

Signatures of high excitation (\heii, \feiii, \oiii lines) are present in most of the spectra. 
Low-excitation shock-excited lines are unusually bright in the spectra of ULNXe as well.
We isolate then a group A (HoII X-1, NGC6946 X-1, M101 Po98 and NGC5204 X-1) requiring    
a strong EUV source to ionize oxygen and a group B (HoIX X-1, IC342 X-1, NGC7331 X-1 and M51 X-7) having higher ratios [SII]$\lambda6717,6731$/H$\beta = 2.3-4$ and requiring shock excitation of their nebulae. 
We note that in fact {\it all the ULXNe} studied in the paper show the shock excitation, 
because their ratios [SII]$\lambda6717,6731$/H$\alpha =0.3-1.1$. 

%The group B has highest [SII]$\lambda6717,6731$/H$\alpha$ratios. 
%Considering the group A we may expect that if one takes into account the hydrogen 
%UV ionization in these ULXNe and derives the photoionization contribution in the 
%hydrogen lines, the residual [SII]$\lambda6717,6731$/H$\alpha$ ratios will increase, 
%because the [SII] line intensities are practically independent on UV photoionization (\cite{ngc7331}). 

We may make (preliminary) conclusions:
\begin{itemize}
\item[---] All the ULXNe show signatures of shock excitation, however the local ISM environment can also change the appearance of a ULX nebula.
\item[---] Half of the ULXNe (the group A) require additional strong  EUV continua (ultraluminous UV sources, UUVs) to ionize the nebulae. 
\end{itemize}

However, this isolation of A and B groups may be done only in average. For example, 
IC342~X-1 nebula (the group B) shows that the central parts of the nebulae is  
bright in high-excitation lines. In fact  all the ULXNe except two (M51~X-7 and NGC7331~X-1 
residual emission) show either HeII $\lambda$4686 emission or enhanced \oiii$\lambda$4959,5007 / H$\beta$ 
ratio which indicate the presence of hard ionizing source with the luminosity $\sim 10^{38}$\ergl.
Most likely shock waves, X-ray and EUV ionization act simultaneously in 
all the ULXNe, but the relative importance of the power sources varies significantly.

In some ULXNe radial velocity gradients ($\sim$100\,km/s) were directly detected
(Dunne \etal (2000), Pakull \etal (2006), Fabrika \etal (2006) and references therein). 
We make an additional conclusion:

\begin{itemize}
\item[---] ULXs must produce strong winds and/or jets powering their
  nebulae with $\gtrsim 10^{39}$\ergl. This is consistent with the
suggestion that ULXs are high-mass X-ray binaries with the supercritical accret
of the SS433 type. From the other hand, accreting IMBHs with standard
  disks are unlikely to have jet or wind activity.
\end{itemize}

Some of ULXs are bright ultraviolet sources with hydrogen- and helium-ionizing luminosities 
varying from $\sim 10^{38}$ to $\sim 10^{40}$\ergl. 
This property can be still explained by a population of IMBHs with comparatively high
($\sim 10000 M\sun$) masses or stellar-mass supercritical accretors (SCAD) like SS433. 
Future UV (Galex) observations may help in solving of this task, we predict different
spectral slopes in the UV spectral ranges in the IMBHs/SCADs models.

\begin{acknowledgements}
 This work was supported by the RFBR grants NN 05-02-19710, 04-02-16349, 06-02-16855. 
\end{acknowledgements}

\end{document}